\documentclass[fleqn,usenatbib]{mnras}

\usepackage{newtxtext,newtxmath}
\usepackage[T1]{fontenc}

\DeclareRobustCommand{\VAN}[3]{#2}
\let\VANthebibliography\thebibliography
\def\thebibliography{\DeclareRobustCommand{\VAN}[3]{##3}\VANthebibliography}

\usepackage{graphicx}	% Including figure files
\usepackage{amsmath}	% Advanced maths commands
\usepackage[utf8]{inputenc} %Process and print 'UTF-8' encoded international text (Unicode). 
\usepackage{ulem} %For revised version

\mathchardef\mhyphen="2D

\title[Oblique Alfv{\'e}n waves in a dusty plasma]{Oblique Alfv{\'e}n waves in a stellar wind environment with dust particles charged by inelastic collisions and by photoionization}

\author[L. B. De Toni, R. Gaelzer, L. F. Ziebell]{
L. B. De Toni,$^{1}$\thanks{E-mail: luan.toni@ufrgs.br}
R. Gaelzer$^{1}$\thanks{E-mail: rudi.gaelzer@ufrgs.br}
and L. F. Ziebell$^{1}$\thanks{E-mail: luiz.ziebell@ufrgs.br}
\\
$^{1}$Instituto de F{\i}sica, Universidade Federal do Rio Grande do Sul, CP 15051, 91501-970, Porto Alegre, RS, Brazil
}

\date{This is a pre-copyedited, author-produced PDF of an article accepted for publication in Monthly Notices of the Royal Astronomical Society following peer review.}

\pubyear{2021}

\begin{document}
\label{firstpage}
\pagerange{\pageref{firstpage}--\pageref{lastpage}}
\maketitle

% Abstract of the paper
\begin{abstract}
The characteristics of Alfv{\'e}n waves propagating in a direction oblique to the ambient magnetic field in a stellar wind environment are discussed. A kinetic formulation for a magnetized dusty plasma is adopted considering Maxwellian distributions of electrons and ions, and immobile dust particles electrically charged by absorption of plasma particles and by photoionization. The dispersion relation is numerically solved and the results are compared with situations previously studied where dust particles were not charged by photoionization, which is an important process in a stellar wind of a relatively hot star. We show that the presence of dust causes the shear Alfv{\'e}n waves to present a region of wavenumber values with zero frequency and that the minimum wavelength for which the mode becomes dispersive again is roughly proportional to the radiation intensity to which the dust grains are exposed. The damping rates of both shear and compressional Alfv{\'e}n waves are observed to decrease with increasing radiation flux, for the parameters considered. For the particular case where both modes present a region with null real frequency when the radiation flux is absent or weak, it is shown that when the radiation flux is sufficiently strong, the photoionization mechanism may cause this region to get smaller or even to vanish, for compressional Alfv{\'e}n waves. In that case, the compressional Alfv{\'e}n waves present non zero frequency for all wavenumber values, while the shear Alfv{\'e}n waves still present null frequency in a certain interval of wavenumber values, which gets smaller with the presence of radiation.
\end{abstract}

% Select between one and six entries from the list of approved keywords.
% Don't make up new ones.

%%%%%%%%%%%%%%%%%%%%%%%%%%%%%%%%%%%%%%%%%%%%%%%%%%

%%%%%%%%%%%%%%%%% BODY OF PAPER %%%%%%%%%%%%%%%%%%

\section{Introduction}
Alfv{\'e}n waves are low-frequency waves ubiquitous in space environments and are of great interest as they play important roles in the heating and transport of energy in laboratory and astrophysical plasmas. They are believed to heat and accelerate stellar winds through the transport of magnetic energy, provide scattering mechanisms for the acceleration of cosmic rays, transfer angular momentum in interstellar molecular clouds, and play important roles in magnetic pulsations in Earth's magnetosphere \citep[][]{cramer2001physics}\@.

These waves have been detected through \textit{in situ} measurements and are known to exist in the solar wind environment, where they can provide the energy flux needed to drive the wind \citep[][]{belcher1971alfvenic,Smith_1995Ulysses,Tomczyk_2007Alfven}\@. For this reason, and the fact that they are generated by oscillations in the magnetic field, Alfv{\'e}n waves are believed to exist in the winds of stars in many regions of the Hertzsprung–Russell diagram.

\citet{hasegawa1976particle} first introduced the kinetic Alfv{\'e}n waves (KAW) to include the effects of finite electron pressure and finite ion gyroradius on the ideal magnetohydrodynamic shear Alfv{\'e}n wave. Since then, KAWs have been invoked in association with various space phenomena, such as auroral currents, particle acceleration in the magnetosphere \citep[][]{hui1992electron} and the dissipation range of the turbulent interplanetary magnetic field \citep[][]{leamon1998observational}\@. \citet{Leamon1999} showed that purely parallel propagating Alfv{\'e}n waves do not adequately describe some of the solar wind characteristics. They argued that the existence of KAWs propagating at large angles in relation to the interplanetary magnetic field are more consistent with the observations.

The transfer of energy between KAWs and plasma particles usually occurs via Landau damping, resulting in the heating of plasma or acceleration of particles along the magnetic field direction. However, it is known that Alfv{\'e}n waves in a dusty plasma present an additional damping mechanism generally stronger than the conventional Landau damping \citep[][]{dejuli_2005,deJuli_2007mode,Gaelzer_2008,Gaelzer_2010,detoni2021}\@. This new feature is due to the variable electrical charge that dust particles acquire by several charging mechanisms such as absorption of charged particles by inelastic collisions, photoionization, secondary electron emission, among others. 

Dust particles can be observed in the solar wind by the presence of the F-corona, the outer part of the Sun's corona, which is illuminated by sunlight scattered or reflected from solid dust particles. The same phenomenon also produces the zodiacal light much farther from the Sun. Observations made during solar eclipses \citep[][]{durst1982two,maihara1985balloon,koutchmy1985properties,Leinert1998} obtained evidence that dust particles can be present inside the solar wind as close as $2\,R_\odot$, where $R_\odot$ is the Sun's radius. On the other hand, recent observations by NASA's Parker Solar Probe obtained evidence of a gradual decrease of F-coronal dust from distances of approximately $19\mhyphen9\,R_\odot$ to the Sun \citep[][]{howard2019near}, which is suggestive of a long-hypothesised dust-free zone \citep[][]{russell1929composition}, a region close to the Sun where dust has been heated and vaporised by the intense sunlight. However, at this point the evidence is still not conclusive and an exact location of dust-free zone depends on grain compositions and grain sizes \citep[see e.g.][]{Krivov1998,Kobayashi2012}\@.

The existence of dust envelopes in other stars also has been observed from infrared emission and absorption characteristics. An interesting phenomenon related to circumstellar dust is observed in R Coronae Borealis (RCB) stars, an exotic group of carbon-rich supergiants that are known for their spectacular declines in brightness (up to $8$ magnitudes) in visible light at irregular intervals, this fading is less pronounced at longer wavelengths. This sudden drop in brightness may be caused by a rapid condensation of carbon-rich dust, resulting in much of the star's light being blocked. \citet{Clayton1992}, through observations of RCB stars, argue that dust grains must be formed in a region between $1.5\mhyphen2$ stellar radii. 

The model of dust-driven winds in carbon stars derived by \citet{gail_sedlmayr_2014} assumes that the dust formation region starts around $1.4$ stellar radii. Furthermore, to explain the observed flux characteristics of several late-type stars, \citet{Danchi1994} modeled dust shells with inner radii as close as $1.7$ stellar radii to the stars. These observations evidence the presence of dust grains in carbon-rich stars in a region smaller than $2$ stellar radii.

Observations of several carbon-rich stars show that dust-to-gas mass density ratio around these stars has typical values of $10^{-3}$ to $10^{-4}$ \citep[][]{Knapp_1985,Bergeat_2005}. The dust that populate these stars is mostly composed of carbon \citep[][]{nanni2021dust} and can vary in sizes, presenting radii from about $10^{-4}$ to $10^{-7}\,\text{cm}$ \citep[][]{kruger1997two}.

Some model calculations for carbon-rich asymptotic giant branch star winds which combine hydrodynamics with radiation pressure on dust grains, time-dependent dust formation, and radiative transfer, show that dust particles play an important role in the observed flow patterns \citep[][]{Sandin_2003,Woitke_2006,Boulangier_2018}. These models consider stars with effective temperatures of $2400\mhyphen2600\,\text{K}$ with dust formation starting around $1.5$ stellar radii. In this region, it is possible to observe a gas density of $10^{-8}\mhyphen10^{-12}\,\text{g}/\text{cm}^3$ and temperatures ranging from around $10^{3}$ to over $10^{4}\,\text{K}$. Interestingly, an axi-symmetric (2D) model shows that these winds' flow patterns are far away from being spherically symmetric \citep[][]{Woitke_2006}, revealing a more complex picture of wind acceleration than other 1D models.

\citet{Falceta_Goncalves_2002} showed that, in addition to the radiation pressure, the inclusion of a new strong damping mechanism caused by the presence of dust particles in the model of \citet{jatenco1989effect} for mass-loss in late-type stars results in an acceleration mechanism of stellar winds more consistent with observations. For this reason, and the already mentioned effects of Alfv{\'e}n waves in several space environments, it is of paramount interest to better understand the effects of dust particles in the propagation and damping of these waves.

\citet{Gaelzer_2008} showed that obliquely propagating Alfv{\'e}n waves are affected by the presence of dust particles charged solely by absorption of plasma particles through inelastic collisions. The results imply that the long wavelength dynamics of Alfvén waves in a dusty plasma is completely different from the usual behaviour observed in a conventional plasma; in the large wavelength limit both the dispersion and the absorption of KAWs are substantially modified.

In this work we will study the modifications that the photoionization process brings to the propagation and absorption of KAWs within the kinetic theory of plasmas. The inclusion of this additional dust charging mechanism in the theory is of great importance when one studies environments where dust particles are exposed to electromagnetic radiation, such as stellar winds. For this, we make use of the formalism developed by \citet{galvao_ziebell2012} for the kinetic theory of magnetized dusty plasma, considering that dust grains are charged by absorption of particles and by photoionization.

Recently, \citet{detoni2021} analysed the effects of the photoemission process within this new formalism for purely parallel propagating Alfv{\'e}n waves in a stellar wind coming from a carbon-rich star. The results show that the coupling between the whistler and ion-cyclotron modes is greatly modified in the large wavelength region once dust particles present null or positive electrical charge, which is only possible when photoemission of electrons by dust particles is considered in the formalism, since the process of absorption of plasma particles tends to negatively charge the dust grains. 

Also, it was shown that the presence of photoionization may greatly modify the damping rates seen in both whistler and ion cyclotron modes since this damping mechanism is related to the inelastic collision frequency, which depends on the dust equilibrium electrical charge. These results are strong indication that the photoionization mechanism will also have significant role in the dispersion and damping of oblique Alfv{\'e}n waves.

As we will show, by applying this formalism to obliquely propagating Alfv{\'e}n waves we notice that, within the parameters considered in this work, the photoionization process tends to diminish the absorption rate of both shear and compressional Alfv{\'e}n modes, and to reduce the region of non propagation of shear Alfv{\'e}n waves. We also see that for the particular set of parameters where all modes are non propagating, the presence of radiation will modify this condition, causing the compressional Alfv{\'e}n modes to present their usual dispersive properties again.

The plan of this paper is as follows. In Section \ref{sec:the_model}, we discuss the basic properties regarding the dusty plasma model employed. Section \ref{sec:disperion_relation} presents the dispersion relation for obliquely propagating Alfv{\'e}n waves with Maxwellian distributions for plasma particles. In Section \ref{sec:numerical_results}, we present some numerical results considering parameters typically found in stellar winds of carbon-rich stars. Finally, the conclusions are presented in Section \ref{sec:conclusions}.

\section{The dusty plasma model} \label{sec:the_model}
In our formulation we consider a homogeneous dusty plasma embedded in an ambient magnetic field $\mathbfit{B}_{0}=B_{0}\mathbfit{e}_{z}$ and exposed to anisotropic radiation. This plasma is composed by electrons, protons and spherical dust particles with constant radius $a$ and variable charge $q_\mathrm{d}$, which originates from inelastic collisions between dust particles and plasma particles of species $\beta$, and from emission of electrons by photoionization.

Dust particles are assumed to be immobile because their mass $m_\mathrm{d}$ is much larger than the plasma particles' masses $m_\beta$. Consequently, this model is restricted to wave frequencies much higher than the characteristic dust frequencies, thereby excluding the modes that can arise from the dust dynamics. That is, we consider the regime in which $\omega\gg\max\left(\omega_\mathrm{d},|\Omega_\mathrm{d}|\right)$, where $\omega_\mathrm{d}$ and $\Omega_\mathrm{d}$ are, respectively, the plasma and cyclotron frequencies of the dust particles.

This model assumes a single grain size, which enable us to derive a relatively simple dispersion relation, as we will see. However, space environments most probably will present populations of particles with different sizes. The dust radii are often described by a distribution function. For example, the interstellar grains can be modelled by a power law distribution, with radii ranging from $0.005\,\mu\text{m}$ to $1\,\mu\text{m}$ \citep[][]{mathis1977size}. Meanwhile, detailed modelling of the formation and growth of dust in carbon-rich stars suggests that the dust density distribution in this environment has a maximum at some given dust size, with decreasing density for smaller and larger grains \citep[][]{Dominik_1989,Hoefner_1992}. \citet{galvao_ziebell2012} proposed the inclusion of a discrete distribution of grain sizes in the formalism for a magnetized dusty plasma using kinetic theory. In future works we intend to make use of this derivation to include a continuous distribution function of grain sizes in the model.

In the presence of dust, the equilibrium number densities of electrons and ions are no longer the same since some of these particles will be absorbed by dust grains and electrons will be emitted by photoionization. Therefore, the quasi-neutrality condition in a dusty plasma is expressed as
\begin{equation}
    \sum_{\beta}n_{\beta0}q_{\beta}+q_\mathrm{d0}n_{d0}=0,
    \label{eq:quasineutrality}
\end{equation}
where $n_{\beta0}$ and $n_{d0}$ are, respectively the equilibrium density of the plasma particles of species $\beta$ and the dust particles, $q_{\beta}$ is the charge of species $\beta$, and $q_\mathrm{d0}=Z_\mathrm{d} e$ is the equilibrium dust charge where $Z_\mathrm{d}$ is the equilibrium dust charge number and $e$ is the elementary charge.

To evaluate the equilibrium dust charge we use the condition of zero surface current
\begin{equation}
    I_{0}(q_\mathrm{d0})=\sum_{\beta}I_{\beta0}(q_\mathrm{d0})+I_\mathrm{p}(q_\mathrm{d0})=0,
\end{equation}
where $I_{0}$ is the total charging current over the surface of a dust grain, $I_{\beta0}$ is the current due to absorption of plasma particles of species $\beta$ in the equilibrium, and $I_\mathrm{p}$ is the photoemission current.

The absorption current is caused by inelastic collisions of plasma particles with dust particles. It is described using the orbital motion limited (OML) theory \citep[see e.g.][]{Allen_1992,Tsytovich_1997}, which neglects the presence of a magnetic field in the charging process. This approximation is valid for weakly magnetized plasmas where $a\ll r_{Le}$, i.e., when the dust particle radius is much smaller than the electron Larmor radius, as shown in a numerical calculation performed by \citet{Chang_1993}\@. 

However, for a strong magnetic field where the dust radius is comparable or grater than the electron Larmor radius, the dust charging process is modified and the dust charge could reduce significantly \citep[see e.g.][]{salimullah2003dust,Kodanova+2019}\@. This effect happens because the orbits of plasma particles are confined to one dimension along the magnetic field lines, making the dusty plasma anisotropic. 

For the values of parameters used in this work, the relation $a\ll r_{Le}$ is always satisfied, making it possible to use the OML theory in our formulation. Therefore, we can express the absorption current as \citep[][]{dejuli_schneider_1998}
\begin{equation}
    I_{\beta0}(q_\mathrm{d})=\pi a^{2}q_{\beta}\int \mathrm{d}^{3}p\left(1-\frac{C_\beta}{p^{2}}\right)H\left(1-\frac{C_\beta}{p^{2}}\right)\frac{p}{m_{\beta}}f_{\beta0},
    \label{eq:I_beta0_root}
\end{equation}
where
\begin{equation}
    C_\beta \equiv \frac{2 q_\mathrm{d} q_\beta m_\beta }{a},    
\end{equation}
and $p$ is the momentum of the plasma particles, $f_{\beta0}$ is the distribution function of species $\beta$ in equilibrium and $H(x)$ is the Heaviside function. Assuming Maxwellian distribution for the plasma particles, equation~\eqref{eq:I_beta0_root} can be evaluated for electrons ($\beta=e$), resulting
\begin{equation}
    I_{e0}(q_\mathrm{d})=-2\sqrt{2\pi}a^{2}en_{e0}v_{T\!e}
    \begin{cases}
        \exp\left(\frac{q_\mathrm{d}e}{ak_\mathrm{B}T_{e}}\right), &q_\mathrm{d}<0\\
        \left(1+\frac{q_\mathrm{d}e}{ak_\mathrm{B}T_{e}}\right), &q_\mathrm{d}\geq0
    \end{cases},
\end{equation}
and for protons ($\beta=i$), resulting
\begin{equation}
    I_{i0}(q_\mathrm{d})=2\sqrt{2\pi}a^{2}en_{i0}v_{T\!i}
    \begin{cases}
        \left(1-\frac{q_\mathrm{d}e}{ak_\mathrm{B}T_{i}}\right), &q_\mathrm{d}\leq0\\
        \exp\left(-\frac{q_\mathrm{d}e}{ak_\mathrm{B}T_{i}}\right), &q_\mathrm{d}>0
    \end{cases},
\end{equation}
where $k_\mathrm{B}$ is the Boltzmann constant, and $v_{T\!\beta}=(k_\mathrm{B}T_{\beta}/m_{\beta})^{1/2}$ and $T_{\beta}$ are, respectively, the thermal velocity and temperature of the plasma particles of species $\beta$\@.

The model that describes the photoelectric emission by dust particles assumes that the number of electrons emitted by unit area by unit time is proportional to the intensity of radiation, with the distribution of momenta of the electrons in the material obeying the Fermi-Dirac statistics. Electrons at the surface of the dust grains have a certain probability $\beta(\nu)$ to absorb the incoming radiation of frequency $\nu$. When the absorption occurs, these electrons can be emitted if the energy of the radiation is greater than the work function $\phi$ of the material of the grain. For the case of a positively charged dust particle, the energy of the emitted electron must overcome the electrostatic attraction by the grain, otherwise it will be reabsorbed by the dust particle.

For the case in which radiation is unidirectional, featuring a continuous spectrum, and propagating in parallel with the ambient magnetic field, the photoelectric current for a spherical dust grain with charge $q_\mathrm{d}$ uniformly distributed over its surface can be written as \citep{galvao_ziebell2012,detoni2021}
\begin{equation}
\begin{alignedat}{2} 
    & I_\mathrm{p}=e\pi a^{2}\int_{\nu_{0}}^{\nu_\mathrm{m}}S_{a}\chi(\nu)\Lambda(\nu) \mathrm{d}\nu, &q_\mathrm{d}\leq0\,,\\
    & I_\mathrm{p}=e\pi a^{2}\int_{\nu_{0}}^{\nu_\mathrm{m}}S_{a}\frac{\Psi(\xi,q_\mathrm{d})}{\Phi(\xi)}\chi(\nu)\Lambda(\nu) \mathrm{d}\nu, \quad&q_\mathrm{d}>0\,,
\end{alignedat}
\label{eq:I_p_continuous}
\end{equation}
where $\chi(\nu)$ is the photoelectric efficiency of the dust material and can be written as \citep[][]{Spitzer_1948}
\begin{equation}
    \chi(\nu)=
    \begin{cases}
        0, &\nu<\nu_{0}\\
        \frac{729}{16}\left(\frac{\nu_{0}}{\nu}\right)^{4}\left(1-\frac{\nu_{0}}{\nu}\right)\chi_\mathrm{m}, &\nu>\nu_{0}
    \end{cases},
\end{equation}
where $\nu_{0}=\phi/h$ is the threshold frequency, $h$ is the Planck constant, and $\chi_\mathrm{m}$ is the maximum value of the photoelectric efficiency. 

Other terms in equation~\eqref{eq:I_p_continuous} are as follows. $\Lambda(\nu)\mathrm{d}\nu$ is the number of photons with frequency between $\nu$ and $\nu+\mathrm{d}\nu$ incident per unit time per unit area, $\nu_\mathrm{m}$ is the upper limit of the spectrum defined by either $\chi(\nu_\mathrm{m})\approx0$ or $\Lambda(\nu_\mathrm{m})\approx0$, and $S_{a}=S_{e}-S_{s}$ where $S_{e}$ and $S_{s}$ are, respectively, the extinction and scattering coefficients, accordingly with Mie theory \citep[see e.g.][]{Sodha_2011}. We consider $S_{a}=1$ which is a fair approximation when $2\pi a/\lambda\geq10$, where $\lambda$ is the wavelength of the radiation. This condition is always satisfied within the parameters used in this work.

We also have the functions
\begin{equation}
    \Phi(\xi)=\int_{0}^{\exp\xi} \frac{\ln(1+\Omega)}{\Omega} \mathrm{d}\Omega=-\text{Li}_{2}(-\exp\xi),
\end{equation}
\begin{equation}
\begin{aligned}
    \Psi(\xi,q_\mathrm{d})=&\frac{eq_\mathrm{d}}{ak_\mathrm{B}T_\mathrm{d}}\ln\left[1+\exp\left(\xi-\frac{eq_\mathrm{d}}{ak_\mathrm{B}T_\mathrm{d}}\right)\right]\\
    &+\Phi\left(\xi-\frac{eq_\mathrm{d}}{ak_\mathrm{B}T_\mathrm{d}}\right),
\end{aligned}
\end{equation}
where $\text{Li}_{2}$ is the polylogarithm function of order $2$, $T_\mathrm{d}$ is the dust temperature and
\begin{equation}
    \xi=\frac{1}{k_\mathrm{B}T_\mathrm{d}}(h\nu-\phi)
\end{equation}
with $\phi$ being the work function of the material.

\section{Dispersion relation for oblique Alfv{\'e}n waves} \label{sec:disperion_relation}
The dielectric tensor for a homogeneous magnetized dusty plasma, with immobile dust particles charged by absorption of plasma particles and by photoionization can be expressed as
\begin{equation}
    \epsilon_{ij}=\epsilon_{ij}^\mathrm{C}+\epsilon_{ij}^\mathrm{A}+\epsilon_{ij}^\mathrm{P},
    \label{eq:dielectric_tensor}
\end{equation} for $\left\{ i,j\right\} =\left\{ x,y,z\right\} $, where the terms $\epsilon_{ij}^\mathrm{C}$ refer to the components which are formally identical to those of a conventional (dustless) plasma, whilst the terms $\epsilon_{ij}^\mathrm{A}$ and $\epsilon_{ij}^\mathrm{P}$ are entirely due to the presence of dust particles, the former is related to the absorption of plasma particles by the dust and the latter is related to the photoionization process. Explicit expressions for these terms can be found, e.g., in \citet{dejuli_schneider_1998} and \citet{galvao_ziebell2012}\@.

The dispersion relation considering a magnetic field along the $z$ direction and a wave vector in an oblique direction lying in the $xz$ plane, i.e., $\mathbfit{k}=k_{\perp}\mathbfit{e}_{x}+k_{\parallel}\mathbfit{e}_{z}$, is given by
\begin{equation}
    \det\begin{pmatrix}\epsilon_{xx}-N_{\text{\ensuremath{\parallel}}}^{2} & \epsilon_{xy} & \epsilon_{xz}+N_{\text{\ensuremath{\parallel}}}N_{\text{\ensuremath{\perp}}}\\
    \epsilon_{yx} & \epsilon_{yy}-N^{2} & \epsilon_{yz}\\
    \epsilon_{zx}+N_{\text{\ensuremath{\parallel}}}N_{\text{\ensuremath{\perp}}} & \epsilon_{zy} & \epsilon_{zz}-N_{\text{\ensuremath{\perp}}}^{2}
    \end{pmatrix}=0,
    \label{eq:Dispersion-equation}
\end{equation}
where $\mathbfit{N}=N_{\perp}\mathbfit{e}_{x}+N_{\parallel}\mathbfit{e}_{z}=\mathbfit{k}c/\omega$ is the refractive index, $\omega$ is the angular frequency and $c$ is the speed of light in vacuum.

As a first approach to the problem, we consider only the `conventional' part of the dielectric tensor, not including the terms that appear due to the dust charging processes in equation~\eqref{eq:dielectric_tensor}.

For Maxwellian distribution of electrons and ions, the conventional part of the dielectric tensor is written as \citep[see e.g.][Appendix]{dejuli_2005}
\begin{equation}
\begin{aligned}
    \epsilon_{ij}^C = \delta_{ij} + \sum_{n=-\infty}^{\infty} \sum_{\beta} \frac{\omega_{p\beta}^2}{\omega n_{\beta0}} \int \mathrm{d}^{3}p     \left(\frac{p_\parallel}{p_\perp} \right)^{\delta_{jz}} \frac{\partial f_{\beta0}}{\partial p_{\perp}} \\
    \times \frac{p_{\parallel}^{\delta_{iz}} p_{\perp}^{\delta_{ix}+\delta_{iy}} \Pi_{ij}^{n\beta}}{\omega-n\Omega_\beta-\frac{k_{\parallel}p_{\parallel}}{m_{\beta}}+\mathrm{i}\nu_{\beta d}^{0}(p)},
\end{aligned}
\label{eq:tensor_conventional}
\end{equation}
where $\omega_{p\beta}$ and $\Omega_\beta$ are, respectively, the plasma and cyclotron frequencies of particles of species $\beta$. We also have the tensor
\renewcommand\arraystretch{1.5}
\begin{equation}
    \mathbf{\Pi}^{n\beta} =\left(\begin{array}{ccc}
    \frac{n^2 J_n^2}{b_\beta^2} & i\frac{n J_n' J_n}{b_\beta} & \frac{n J_n^2}{b_\beta^2}\\
    -i\frac{n J_n' J_n}{b_\beta} & J_n'^{2} & -iJ_n' J_n\\
    \frac{n J_n^2}{b_\beta^2} & iJ_n' J_n & J_n^2
    \end{array}\right),
\end{equation}
where $b_\beta=k_\perp p_\perp/m_\beta \Omega_\beta$, $J_n=J_n(b_\beta)$ is the Bessel function of order $n$ and $J_n'$ is its first derivative.

For the evaluation of momentum integrals which appear in equation~\eqref{eq:tensor_conventional} we follow the same procedure from previous works \citep[][]{dejuli_2005,Ziebell_2005,Gaelzer_2008,detoni2021} where the momentum dependent inelastic collision frequency
\begin{align}
   & \nu_{\beta d}^{0}(p)=\frac{\pi a^{2}n_{d0}}{m_{\beta}}\frac{\left(p^{2}-\frac{2m_{\beta}q_{\beta}q_\mathrm{d0}}{a}\right)}{p}H\left(p^{2}-\frac{2m_{\beta}q_{\beta}q_\mathrm{d0}}{a}\right)
\end{align}
is replaced by its average values in momentum space
\begin{equation}
    \nu_{\beta}=\frac{1}{n_{\beta0}}\int \mathrm{d}^{3}p\,\nu_{\beta d}^{0}(p)f_{\beta0}.
\end{equation}
Considering Maxwellian distributions for protons ($\beta=i$) and electrons ($\beta=e$) we obtain
\begin{align}
    & \nu_{i}=2\sqrt{2\pi}a^{2}n_{d0}v_{T\!i}
    \begin{cases}
        \left(1-\frac{eq_\mathrm{d0}}{ak_\mathrm{B}T_{i}}\right), &q_\mathrm{d0}\leq0\\
        \exp\left(-\frac{eq_\mathrm{d0}}{ak_\mathrm{B}T_{i}}\right), &q_\mathrm{d0}>0
    \end{cases},\\
    & \nu_{e}=2\sqrt{2\pi}a^{2}n_{d0}v_{T\!e}
    \begin{cases}
        \exp\left(\frac{eq_\mathrm{d0}}{ak_\mathrm{B}T_{e}}\right), &q_\mathrm{d0}<0\\
        \left(1+\frac{eq_\mathrm{d0}}{ak_\mathrm{B}T_{e}}\right), &q_\mathrm{d0}\geq0
    \end{cases}.
\end{align}

\citet{deJuli_2007mode} showed that in the frequency range of Alfv{\'e}n waves the dispersion relation can be satisfyingly described with the use of the average collision frequency approximation. This enables us to arrive at a relative simple expression for the dispersion relation.

We now introduce the following dimensionless parameters
\begin{equation}
\begin{alignedat}{2}
    &z=\frac{\omega}{\Omega_{i}},\quad \varepsilon=\frac{n_{d0}}{n_{i0}},\quad  u_{\beta}=\frac{v_{T\!\beta}}{v_\mathrm{A}},\quad \chi_{\beta}=\frac{q_\mathrm{d0}q_{\beta}}{ak_\mathrm{B}T_{\beta}},\\
    &\gamma=\frac{\lambda^{2}n_{i0}v_\mathrm{A}}{\Omega_{i}},\quad  \tilde{a}=\frac{a}{\lambda},\quad \lambda=\frac{e^{2}}{k_\mathrm{B}T_{i}},\quad  \mathbfit{q}=\frac{\mathbfit{k}v_\mathrm{A}}{\Omega_{i}},\\
    &\tilde{\nu}_{\beta}=\frac{\nu_{\beta}}{\Omega_{i}},\quad \eta_{\beta}=\frac{\omega_{p\beta}}{\Omega_{i}},\quad  r_{\beta}=\frac{\Omega_{\beta}}{\Omega_{i}},
\end{alignedat}
\end{equation}
where $v_\mathrm{A}$ is the Alfvén velocity,
\begin{equation}
    v_\mathrm{A}^{2}=\frac{B_{0}^{2}}{4\pi n_{i0}m_{i}}.
\end{equation}
The collision frequencies in terms of these quantities are given by
\begin{equation}
    \tilde{\nu}_{\beta}=2\sqrt{2\pi}\varepsilon\gamma\tilde{a}^{2}u_{\beta}
    \begin{cases}
        \left(1+|\chi_{\beta}|\right), &\chi_{\beta}\leq0\\
        \exp\left(-\chi_{\beta}\right), &\chi_{\beta}>0
    \end{cases}.
    \label{eq:colision_freq}
\end{equation}

The dispersion relation is evaluated considering that the studied modes have large wavelength in the perpendicular direction, i.e., $q_\perp\ll1$, and keeping only the $n=-1,0,1$ harmonics in the dielectric components. A more detailed account of this derivation is given by \citet{Gaelzer_2008}, and as a result we have the following expression for the dispersion relation for obliquely propagating waves,
\begin{equation}
    \begin{aligned}
        &\bigg[\bigg(\frac{z^{2}}{\eta_{i}^{2}}+\epsilon_{yy}^{1}-q_{\parallel}^{2}\bigg) \bigg(\frac{z^{2}}{\eta_{i}^{2}}+\epsilon_{xx}^{1}-q_{\parallel}^{2}\bigg) - (\epsilon_{xy}^{1})^{2} \bigg] \bigg(\frac{z^{2}}{\eta_{i}^{2}}+\epsilon_{zz}^{0}\bigg)\\
        &+\Bigg\{\bigg(\frac{z^{2}}{\eta_{i}^{2}}+\epsilon_{zz}^{0}\bigg) \bigg(\frac{z^{2}}{\eta_{i}^{2}}+\epsilon_{xx}^{1}-q_{\parallel}^{2}\bigg) \bigg(\epsilon_{yy}^{0}-1\bigg)\\
        &\quad+\bigg(\frac{z^{2}}{\eta_{i}^{2}}+\epsilon_{yy}^{1}-q_{\parallel}^{2}\bigg) \bigg[ \bigg(\epsilon_{zz}^{1}-1\bigg) \bigg(\frac{z^{2}}{\eta_{i}^{2}}+\epsilon_{xx}^{1}-q_{\parallel}^{2}\bigg) \\
        &\quad- \bigg(\epsilon_{xz}^{1}+q_{\parallel}\bigg)^{2} \bigg] -\bigg[ - 2\epsilon_{xy}^{1}\epsilon_{yz}^{1}\bigg(\epsilon_{xz}^{1}+q_{\parallel}\bigg) \\
        &\quad+(\epsilon_{xy}^{1})^{2}\bigg(\epsilon_{zz}^{1}-1\bigg)+(\epsilon_{yz}^{1})^{2}\bigg(\frac{z^{2}}{\eta_{i}^{2}}+\epsilon_{xx}^{1}-q_{\parallel}^{2}\bigg) \bigg] \Bigg\} q_{\perp}^{2} \\
        &+\Bigg\{\bigg(\epsilon_{yy}^{0}-1\bigg)\bigg[\bigg(\frac{z^{2}}{\eta_{i}^{2}}+\epsilon_{xx}^{1}-q_{\parallel}^{2}\bigg)\bigg(\epsilon_{zz}^{1}-1\bigg)\\
        &\quad-\bigg(\epsilon_{xz}^{1}+q_{\parallel}\bigg)^{2}\bigg]\Bigg\} q_{\perp}^{4} =0,
    \end{aligned}
    \label{eq:disp_rel}
\end{equation}
where $\eta_i=c/v_\mathrm{A}$ and the $\epsilon_{ij}^{0,1}$ tensor components are related to the $\epsilon_{ij}$ components, remembering that these are the `conventional' part of the dielectric tensor, where we dropped the superscript $C$ since we are not including the other terms in equation~\eqref{eq:dielectric_tensor}, as discussed before. Explicit expressions for these components can be found in \citet{Gaelzer_2008}.

\section{Numerical results} \label{sec:numerical_results}
To numerically solve equation~\eqref{eq:disp_rel} we consider parameters typical of stellar winds coming from carbon-rich stars \citep[][]{tsytovich_2004}: $B_{0}=1$\,G, $n_{i0}=10^{9}$\,cm$^{-3}$, $T_{i}=10^4$\,K, $T_{e}=T_{i}$ and $a=10^{-4}$\,cm. These are the same parameters already employed in previous works using a kinetic approach to study Alfv{\'e}n waves in a dusty plasma \citep[][]{dejuli_2005,Ziebell_2005,Gaelzer_2008,detoni2021}\@. We point out that for the adopted parameters the following conditions are satisfied: $v_{T\!e}>v_\mathrm{A}$ and $m_e/m_i \ll \beta < 1$, where $\beta$ is the plasma parameter. Therefore, we can safely classify the studied waves as KAWs \citep[][]{stasiewicz2000small}\@.

This set of parameters is maintained fixed throughout this work in order to focus our analysis mainly on the changes caused by the photoionization process in the propagation and damping of KAWs. Nonetheless, we point out that these values may vary significantly in a stellar environment. 

\begin{figure}
    \centering
    \begin{minipage}[c]{\columnwidth}
        \centering
        \includegraphics[width=\columnwidth]{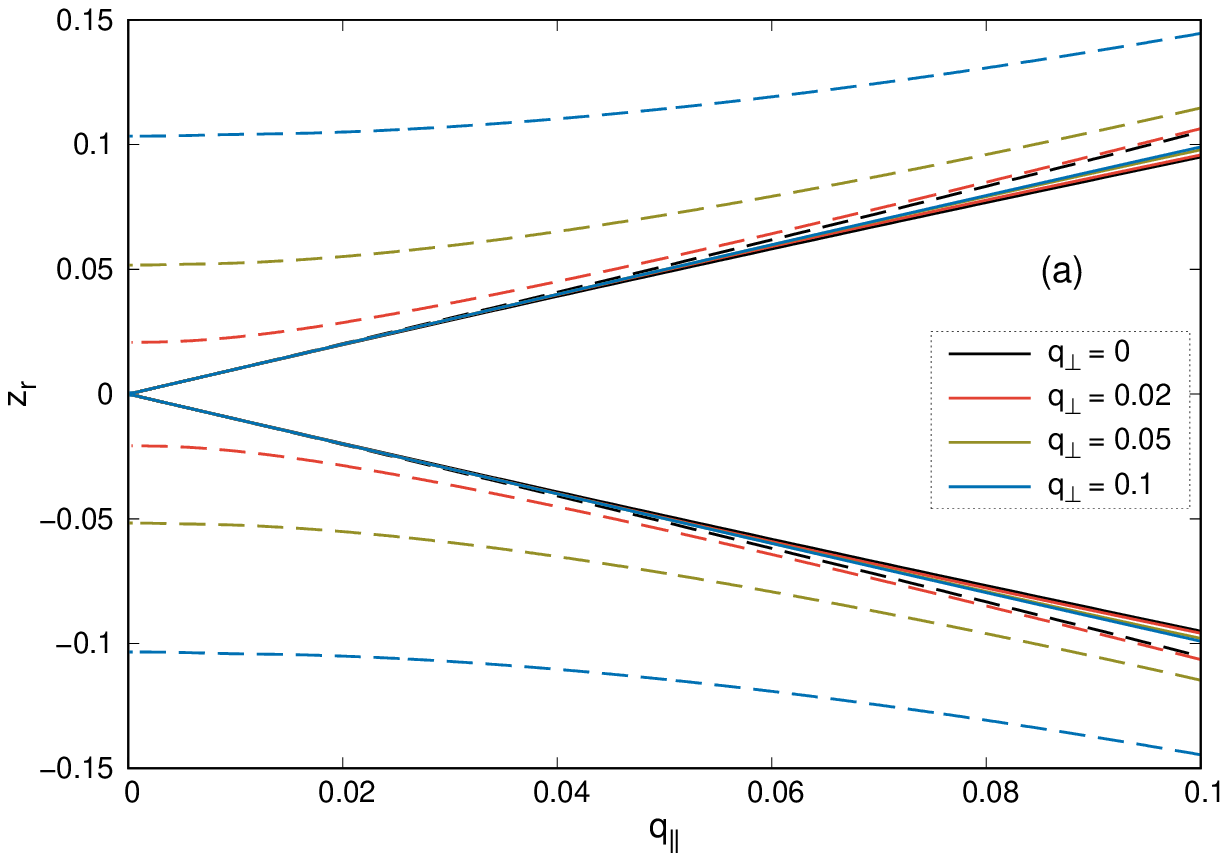}%
    \end{minipage}\vspace{\floatsep}
    \begin{minipage}[c]{\columnwidth}
        \centering
        \includegraphics[width=\columnwidth]{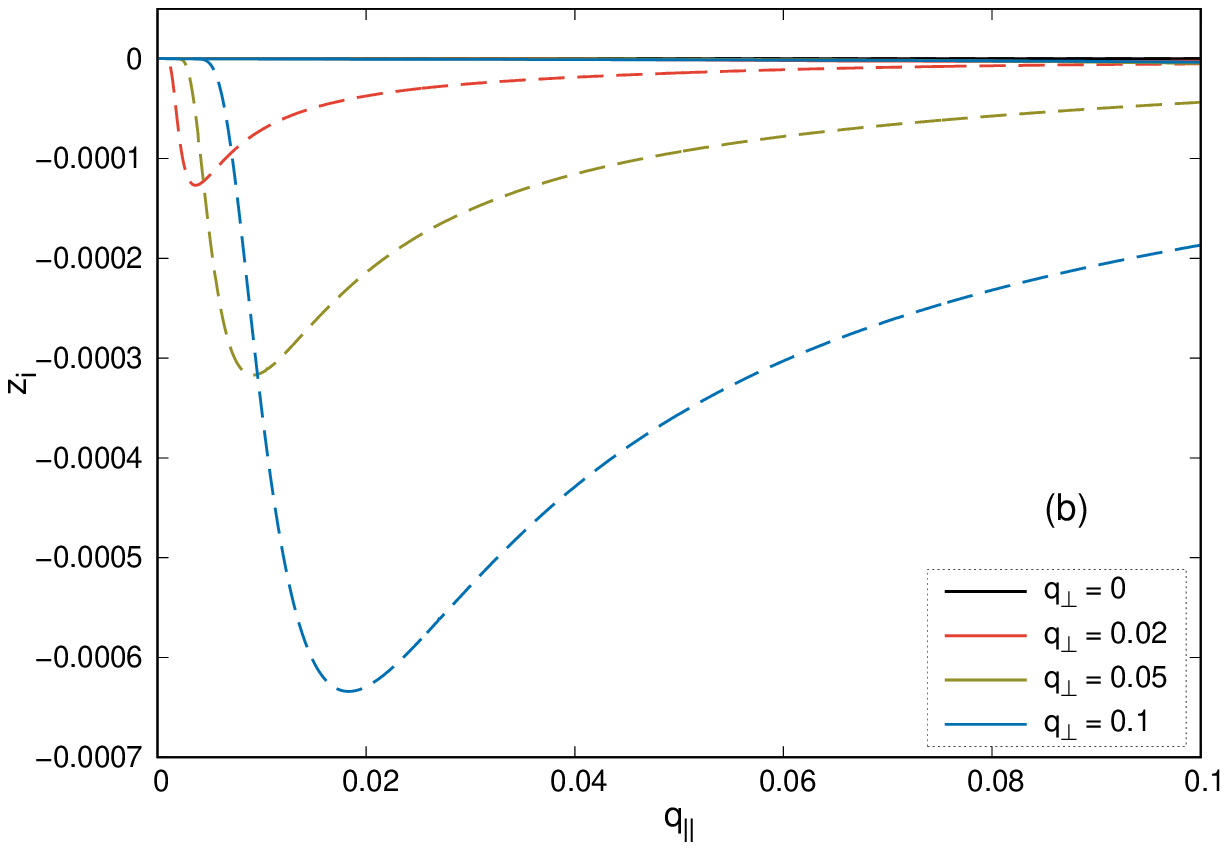}
    \end{minipage}
    \caption{Real (top panel) and imaginary (bottom panel) part of the normalized frequency $z$ as a function of normalized wavenumber in the parallel direction $q_\parallel$ for several values of perpendicular wavenumber $q_\perp$ for a dustless plasma ($\varepsilon=0$). Continuous lines are the shear Alfv{\'e}n modes whilst dashed lines are the compressional Alfv{\'e}n waves.}
    \label{fig:z_q_qperp}
\end{figure}

We consider that the dust particles are exposed to a blackbody radiation spectrum that emanates from the stellar surface. Thus, we express the term related to the photon flux in equation~\eqref{eq:I_p_continuous} as
\begin{equation}
    \Lambda(\nu)\mathrm{d}\nu=\frac{4\pi\nu^{2}}{c^{2}}\left[\exp\left(\frac{h\nu}{k_\mathrm{B}T_\mathrm{s}}\right)-1\right]^{-1}\left(\frac{r_\mathrm{s}}{r_\mathrm{d}}\right)^{2}\mathrm{d}\nu,
    \label{eq:blackbody}
\end{equation}
where $r_\mathrm{s}$ is the radius of the star's radiating surface, $r_\mathrm{d}$ is the mean distance of dust particles from the star, which we consider as $r_\mathrm{d}=2\,r_\mathrm{s}$, unless told otherwise, and $T_\mathrm{s}$ is the stellar surface temperature.

The incoming photon flux may be modified by the extinction of light provoked by dust particles situated between the font of radiation and the region of interest \citep[][]{Massey_2005}. This causes a reddening effect at short wavelengths and may significantly modify the observed stellar spectra when radiation goes through a considerable amount of circumstellar dust \citep[][]{truong2021modeling}. As a first approach, we neglect this effect that would add considerable complexity to the model, and concentrate on the effect of the photoionization for a given set of parameters. Therefore, we consider that there is no dust between the font and the studied region, an approximation that seems to be more justified by the fact that we assume distances very close to the stellar surface.

Carbon stars may present a wide range of surface temperatures \citep[see e.g.][]{wallerstein_1998_carbonstars} from about $2000$\,K to over $5000$\,K. Particularly, hotter carbon stars with $T_s \geq 3500$\,K are of our interest since these systems present high radiation flux with relatively low plasma temperatures, enabling us to better understand the effects of the photoionization process in the propagation and damping of Alfv{\'e}n waves.

Dust particles in the surroundings of carbon-rich stars are commonly composed of carbon \citep[][]{nanni2021dust}, with work function of $\phi=4.6$\,eV \citep{Sohda_1997_glassycarbon}, maximum photoelectric efficiency of $\chi_\mathrm{m}=0.05$ \citep{Feuerbacher_1972} and dust temperature of $T_\mathrm{d}=300$\,K, which is within the range of temperatures of dust in the inner circumstellar dust shells of carbon stars \citep{gail_sedlmayr_2014}\@.

If we considered an oxygen-rich star like the Sun, it would be more likely to find silicate grains in its surroundings which have smaller photoelectric yield \citep[][]{Feuerbacher_1972}, reducing the photoemission of electrons when compared to carbon grains. Moreover, silicate grains present a higher secondary emission yield \citep[][]{chow1993role} and the solar wind has much higher plasma temperatures, these factors would make the secondary emission of electrons an important charging mechanism of dust particles, which is not considered in our model. Therefore, Alfv{\'e}n waves in carbon-rich stars may present more important effects related to the photoionization process, which is our focus in this work.

We start our study by numerically solving equation~\eqref{eq:disp_rel} for the situation of a dustless plasma ($\varepsilon$=0). Fig.~\ref{fig:z_q_qperp} shows the real (top panel) and imaginary (bottom panel) parts of the normalized wave frequency $z$ as functions of the parallel wavenumber $q_\parallel$. This will be useful to compare and identify the different wave modes when we consider the presence of dust particles. 

The continuous lines are identified as shear/slow Alfv{\'e}n waves for small wavenumber and becomes the ion cyclotron mode for large wavenumber \citep[see][]{Gaelzer_2008}\@. This mode disappears for perpendicular propagation. The dashed lines are identified as compressional/fast Alfv{\'e}n waves for small wavenumber and becomes the whistler mode for large wavenumber. The positive solutions correspond to forward-propagating waves relative to the magnetic field, while the negative solutions are related to backward-propagating waves.

\begin{figure}
    \centering
    \begin{minipage}[c]{\columnwidth}
        \centering
        \includegraphics[width=\columnwidth]{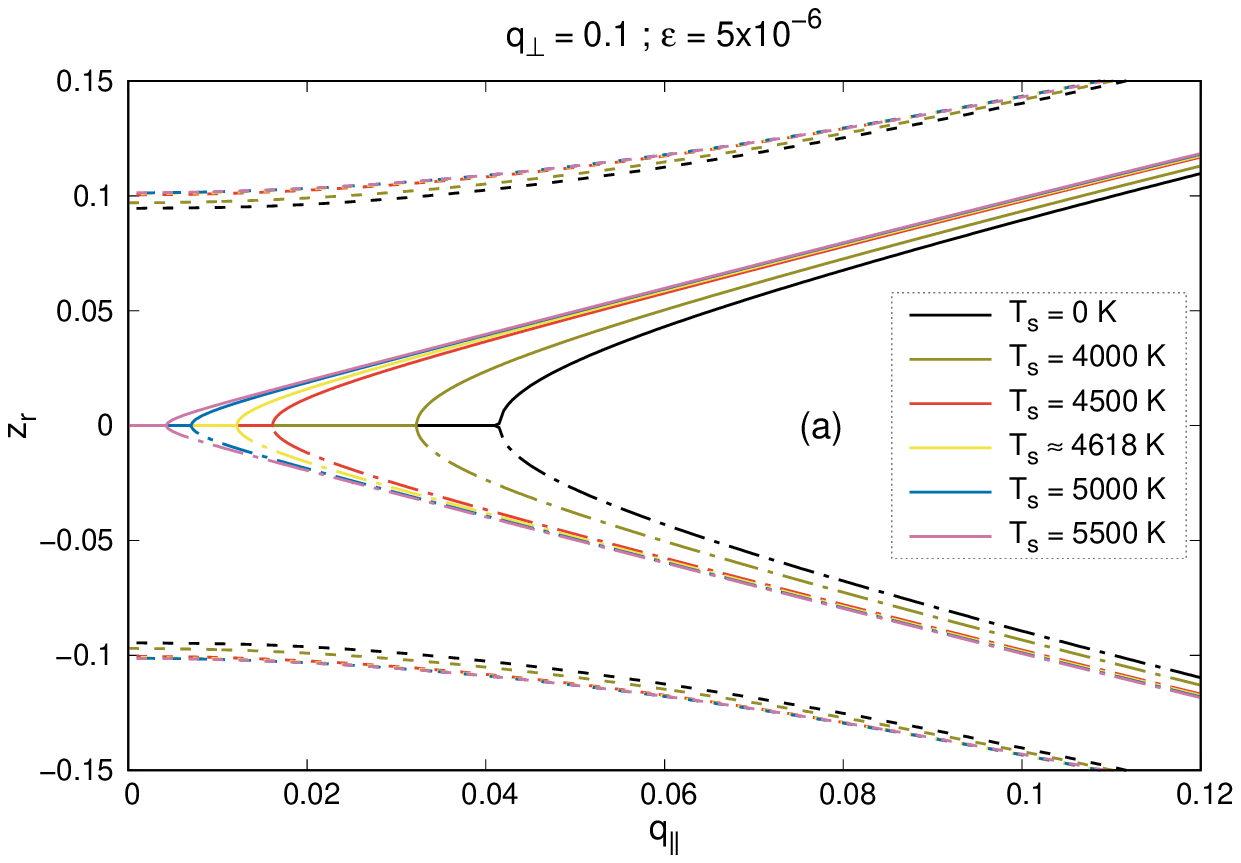}%
    \end{minipage}\vspace{\floatsep}
    \begin{minipage}[c]{\columnwidth}
        \centering
        \includegraphics[width=\columnwidth]{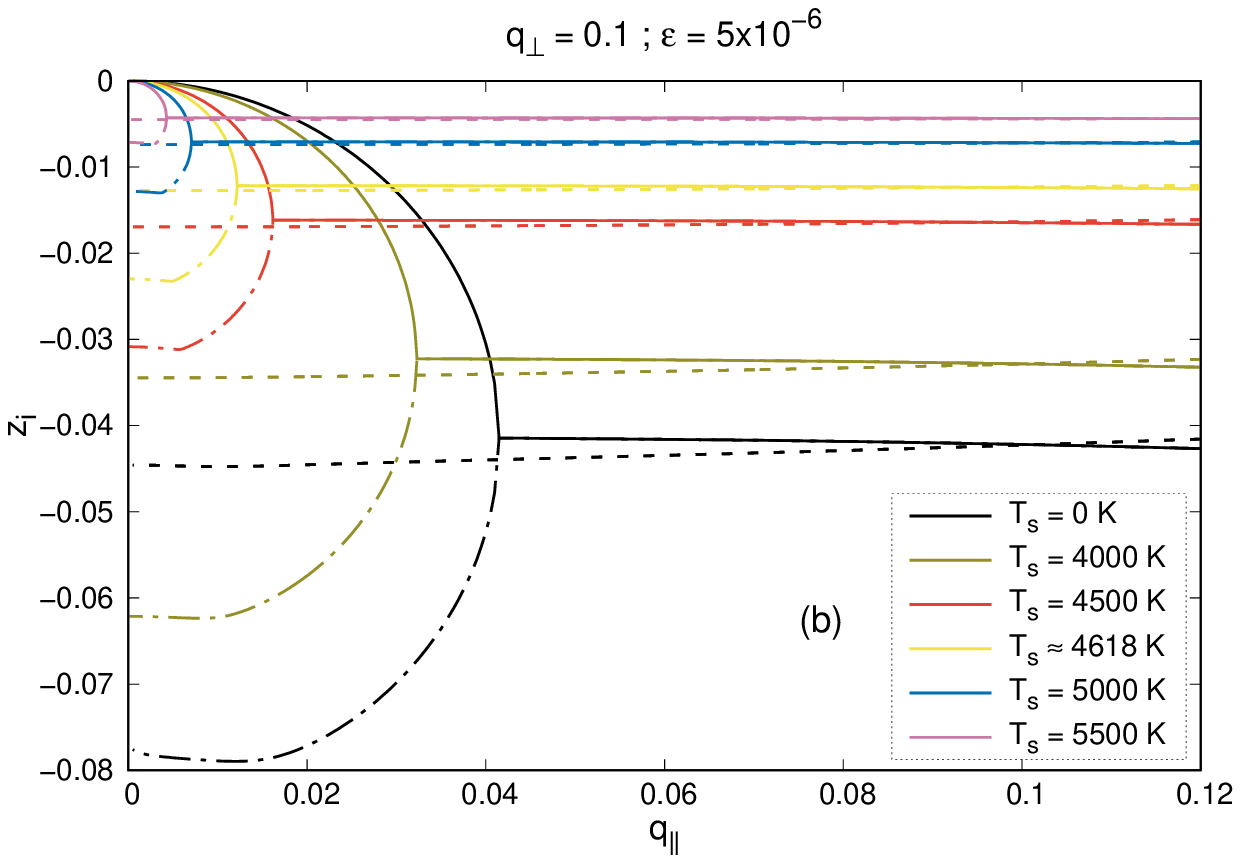}
    \end{minipage}
    \caption{Real (top panel) and imaginary (bottom panel) part of $z$ as a function of $q_\parallel$ for several values of stellar surface temperature $T_s$ and fixed values of $q_\perp=0.1$ and $\varepsilon=5\times10^{-6}$. Dashed lines are the whistler/compressional Alfv{\'e}n waves whilst the continuous and dotted-dashed lines are, respectively, the forward and backward propagating ion cyclotron/shear Alfv{\'e}n modes. The corresponding numerical dust charge numbers are as follows: $Z_\mathrm{d}\simeq-1495$ ($0$\,K); $Z_\mathrm{d}\simeq-1025$ ($4000$\,K); $Z_\mathrm{d}\simeq-205$ ($4500$\,K); $Z_\mathrm{d}=0$ ($4618$\,K); $Z_\mathrm{d}\simeq+349$ ($5000$\,K); and $Z_\mathrm{d}\simeq+700$ ($5500$\,K).}
    \label{fig:z_q_Ts}
\end{figure}

We notice in Fig.~\ref{fig:z_q_qperp}(a) that for parallel propagation the two modes couple for small wavenumber, separating into the whistler and ion cyclotron modes for large values of $q_\parallel$. However, when we consider $q_\perp>0$ these modes decouple for all values of wavenumber since the compressional Alfv{\'e}n waves approach a non zero value of $z_r$ for very small $q_\parallel$. Moreover, we point out that in the presence of dust particles these two modes also may decouple for very small values of wavenumber even for parallel propagation \citep[see e.g.][]{dejuli_2005,Gaelzer_2008,detoni2021}.

Fig.~\ref{fig:z_q_qperp}(b) shows that these modes present no damping rate (given by $z_i$) for parallel propagation ($q_\perp=0$) in a dustless plasma. However, for oblique propagation, these waves may show non zero values for their damping rate, specially the compressional Alfv{\'e}n waves which show significant absorption for large values of $q_\perp$. Moreover, the shear Alfv{\'e}n waves also show higher damping rates for larger $q_\perp$, although these values are quite low in this scale. Here the forward and backward propagating waves of the same mode present similar damping rates.

The effects of dust particles in oblique Alfv{\'e}n waves can be appreciated in Fig.~\ref{fig:z_q_Ts} where the perpendicular wavenumber is fixed at $q_\perp=0.1$ and the ratio of dust to ion densities is set as $\varepsilon=5\times10^{-6}$. We observe in the top panel that the ion cyclotron modes present a region of null $z_r$, a feature that arises with the presence of dust particles. Here we identify the whistler modes as the dashed curves, whilst the forward and backward propagating ion cyclotron modes are the continuous and dotted-dashed curves, respectively. This distinction between different direction of propagation is made because they show distinct damping rates in the non propagating region, as can be seen in the bottom panel.

We notice by the imaginary part of $z$ that the whistler modes show almost constant values of damping rate with increasing $q_\parallel$, a different behaviour of that observed for a dustless plasma in Fig.~\ref{fig:z_q_qperp}. Also, we see that the ion cyclotron modes are now strongly absorbed when compared with the dustless plasma case. The presence of dust particles induces a new damping mechanism in both modes related to the average inelastic collision frequency between them and the plasma particles \citep[see e.g.][]{dejuli_2005,Ziebell_2005,detoni2021}.

The different line colours in Fig.~\ref{fig:z_q_Ts} represent distinct radiation fluxes, given by the stellar surface temperature $T_s$. The variation of the star's temperature will mainly alter the equilibrium charge of dust particles, thus modifying the collision frequencies between dust and plasma particles. The case when radiation is not considered shows a negative dust charge number $Z_\mathrm{d}$, since only the absorption of plasma particles is considered as charging mechanism, which tends to negatively charge the dust grains. 

\begin{figure*}
    \centering
    \begin{minipage}[t]{0.49\textwidth}%
        \centering
        \includegraphics[width=.84\columnwidth]{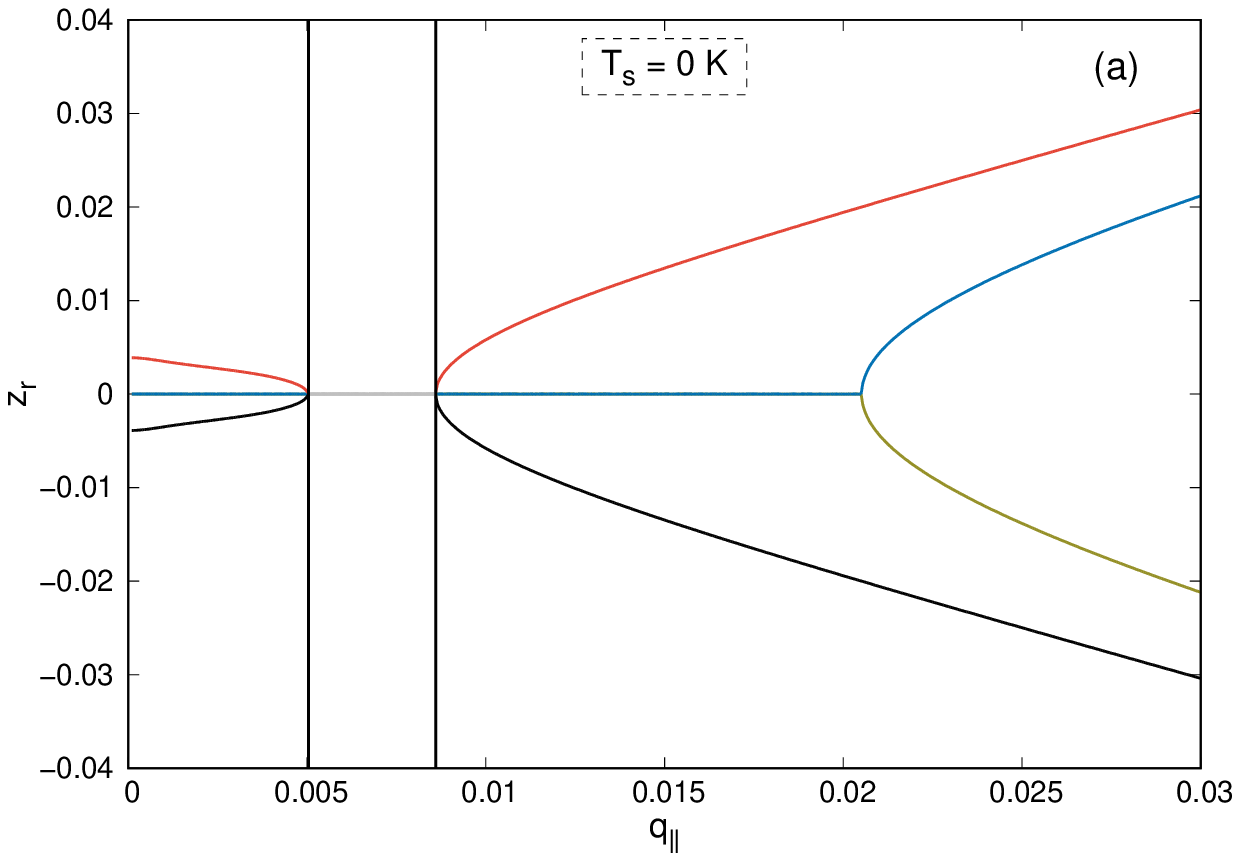}%
    \end{minipage}\hfill{}%
    \begin{minipage}[t]{0.49\textwidth}%
        \centering
        \includegraphics[width=.84\columnwidth]{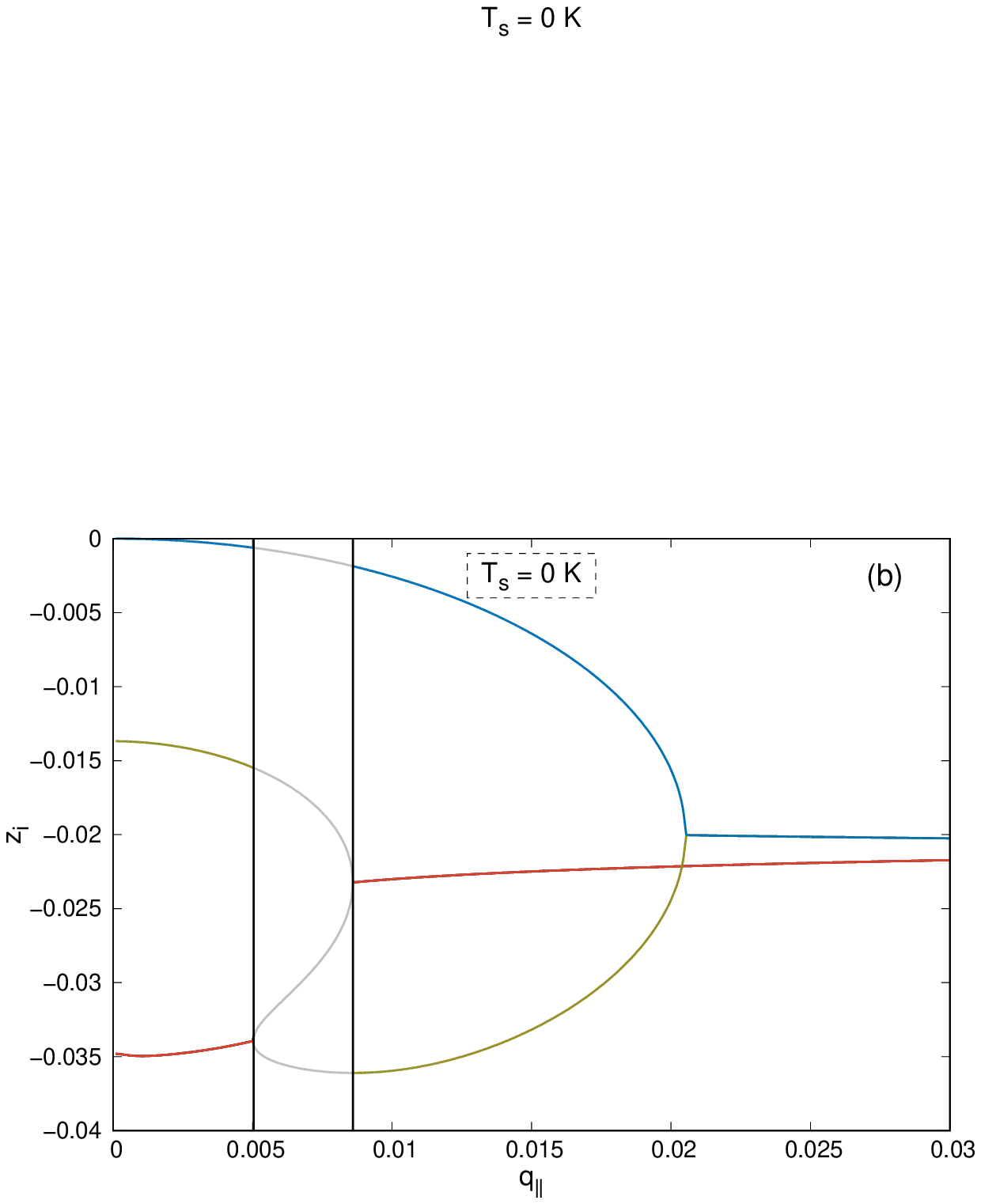}%
    \end{minipage}\vspace{\floatsep}
    \begin{minipage}[t]{0.49\textwidth}%
        \centering
        \includegraphics[width=.84\columnwidth]{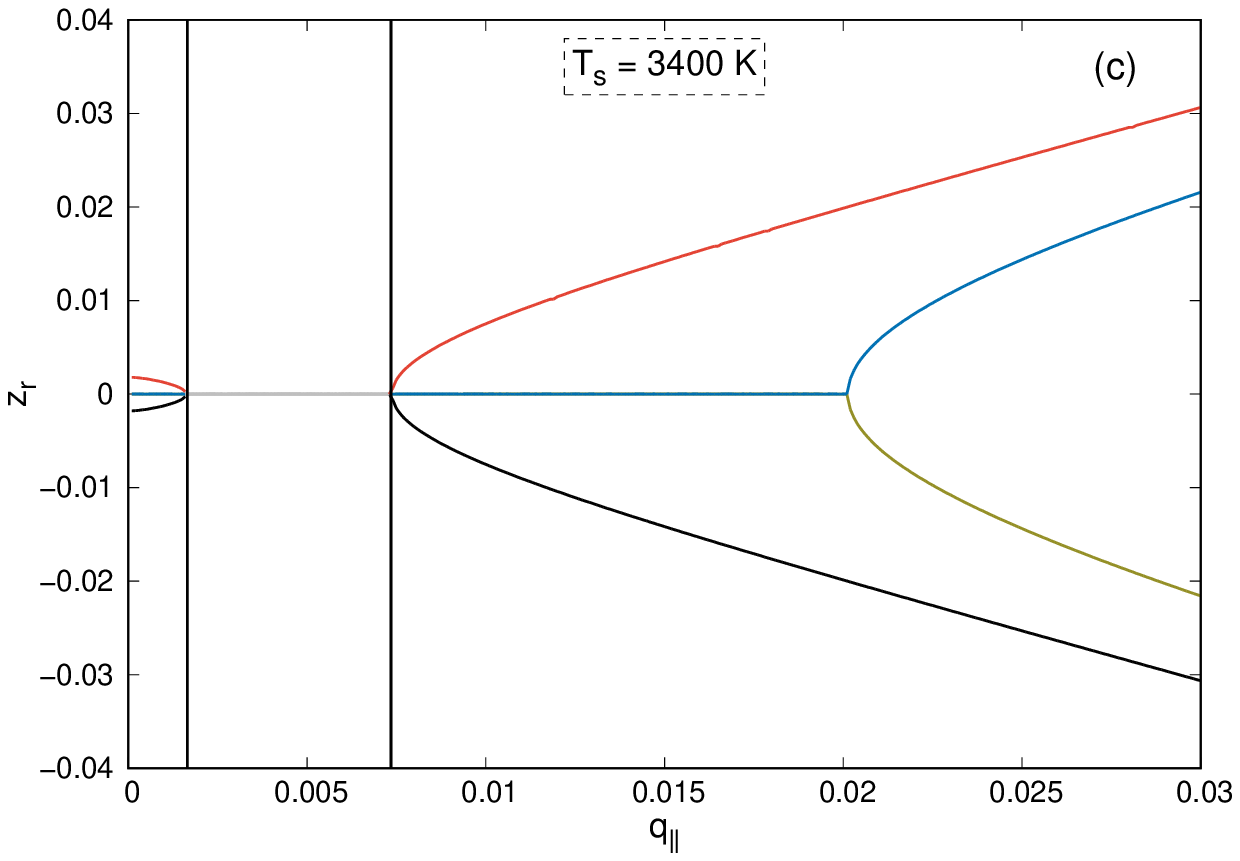}%
    \end{minipage}\hfill{}%
    \begin{minipage}[t]{0.49\textwidth}%
        \centering
        \includegraphics[width=.84\columnwidth]{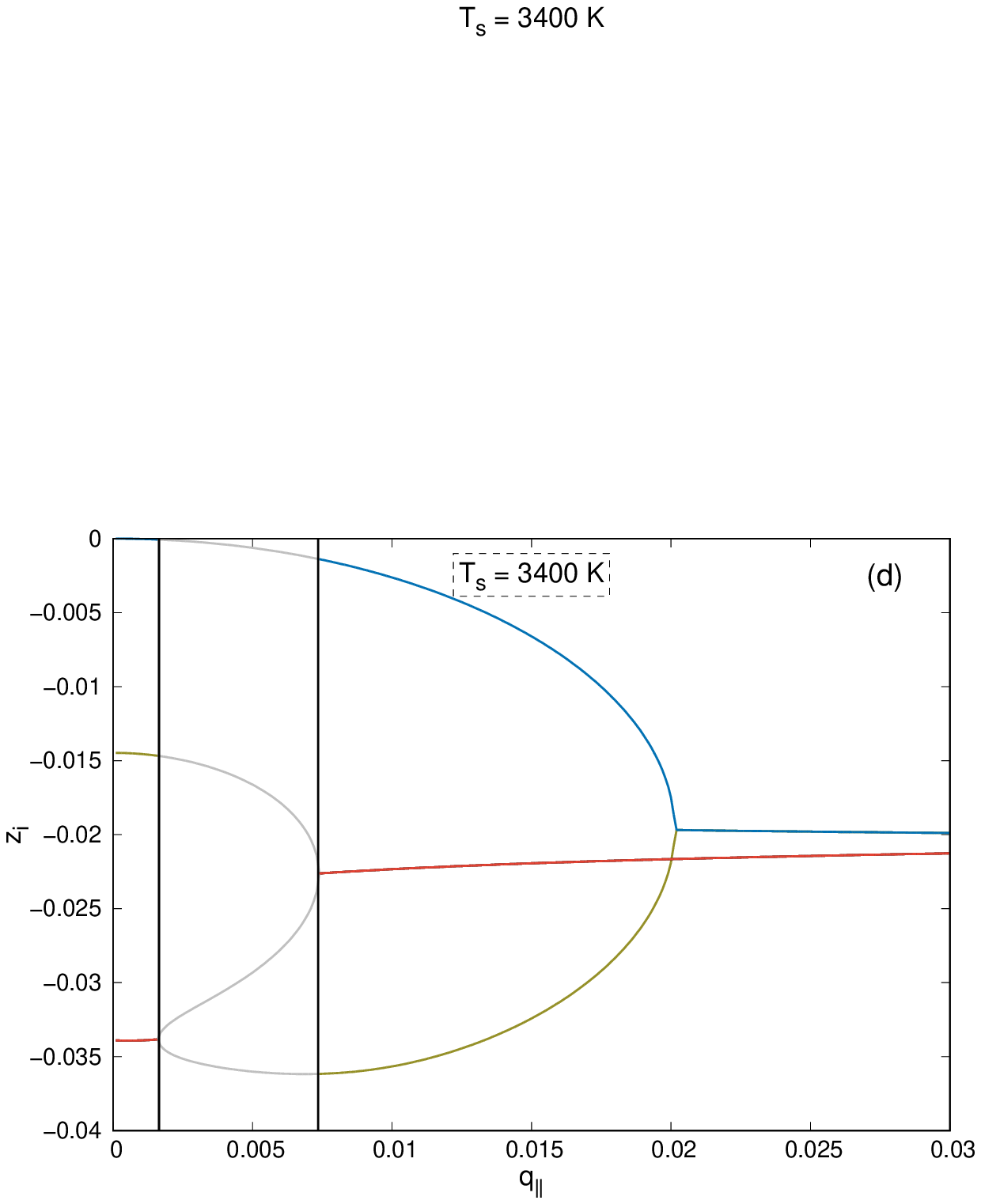}%
    \end{minipage}\vspace{\floatsep}
    \begin{minipage}[t]{0.49\textwidth}%
        \centering
        \includegraphics[width=.84\columnwidth]{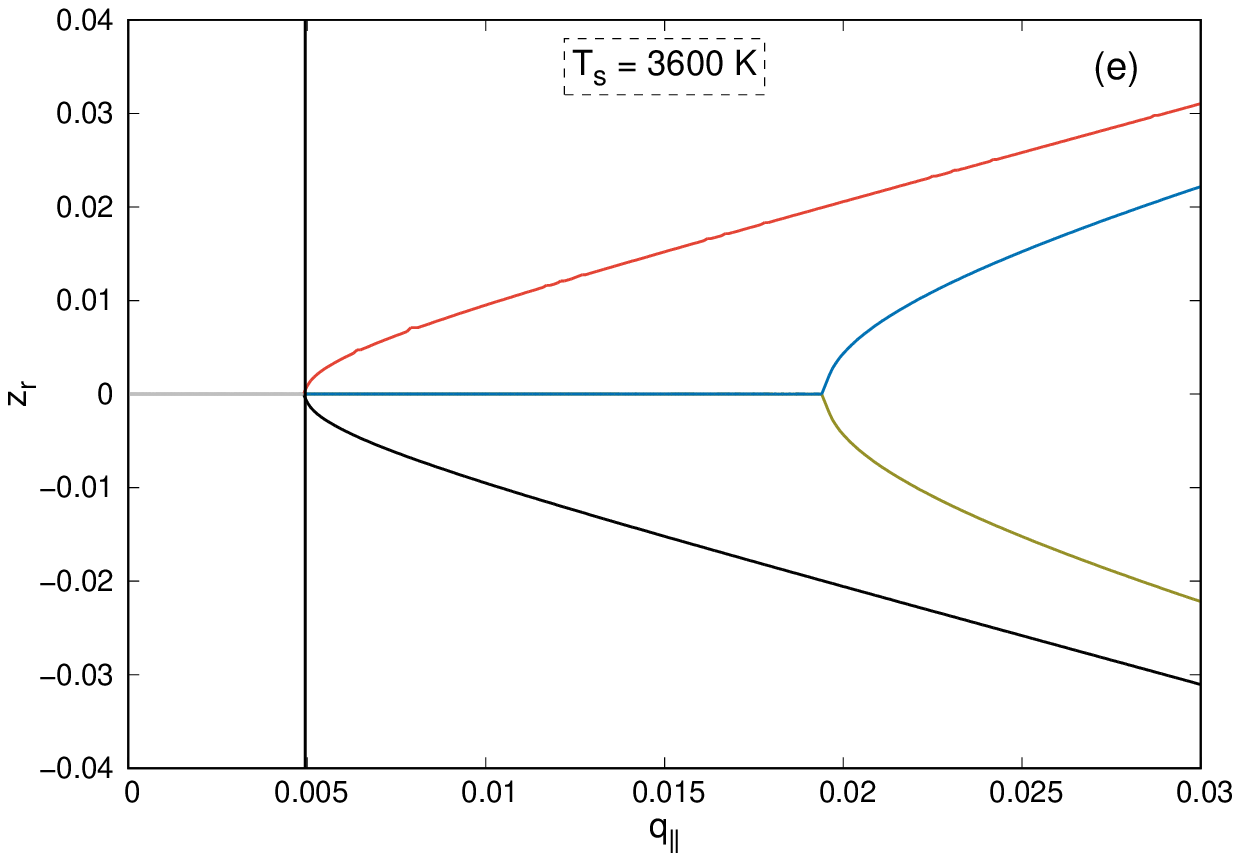}%
    \end{minipage}\hfill{}%
    \begin{minipage}[t]{0.49\textwidth}%
        \centering
        \includegraphics[width=.84\columnwidth]{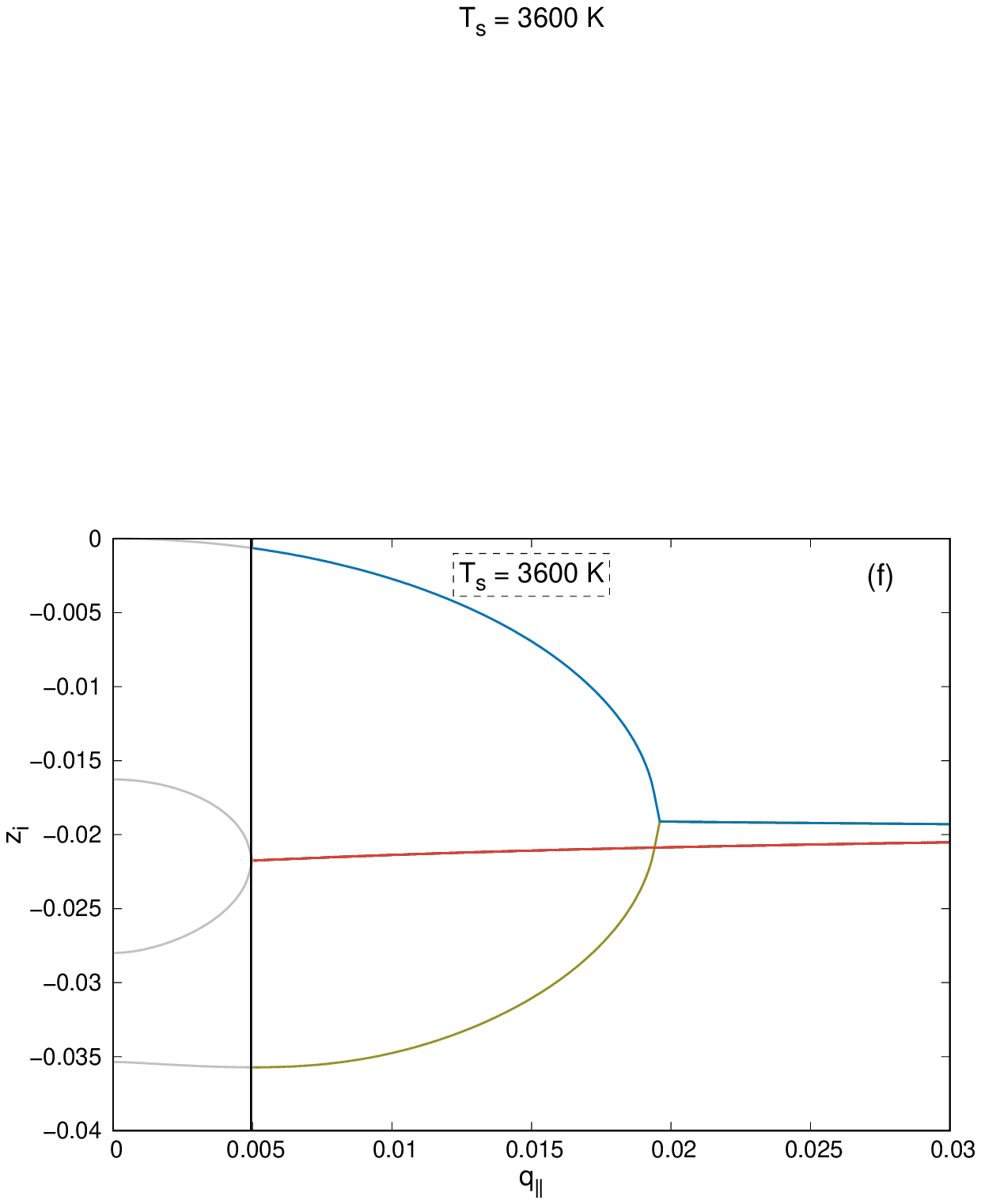}%
    \end{minipage}\vspace{\floatsep}
    \begin{minipage}[t]{0.49\textwidth}%
        \centering
        \includegraphics[width=.84\columnwidth]{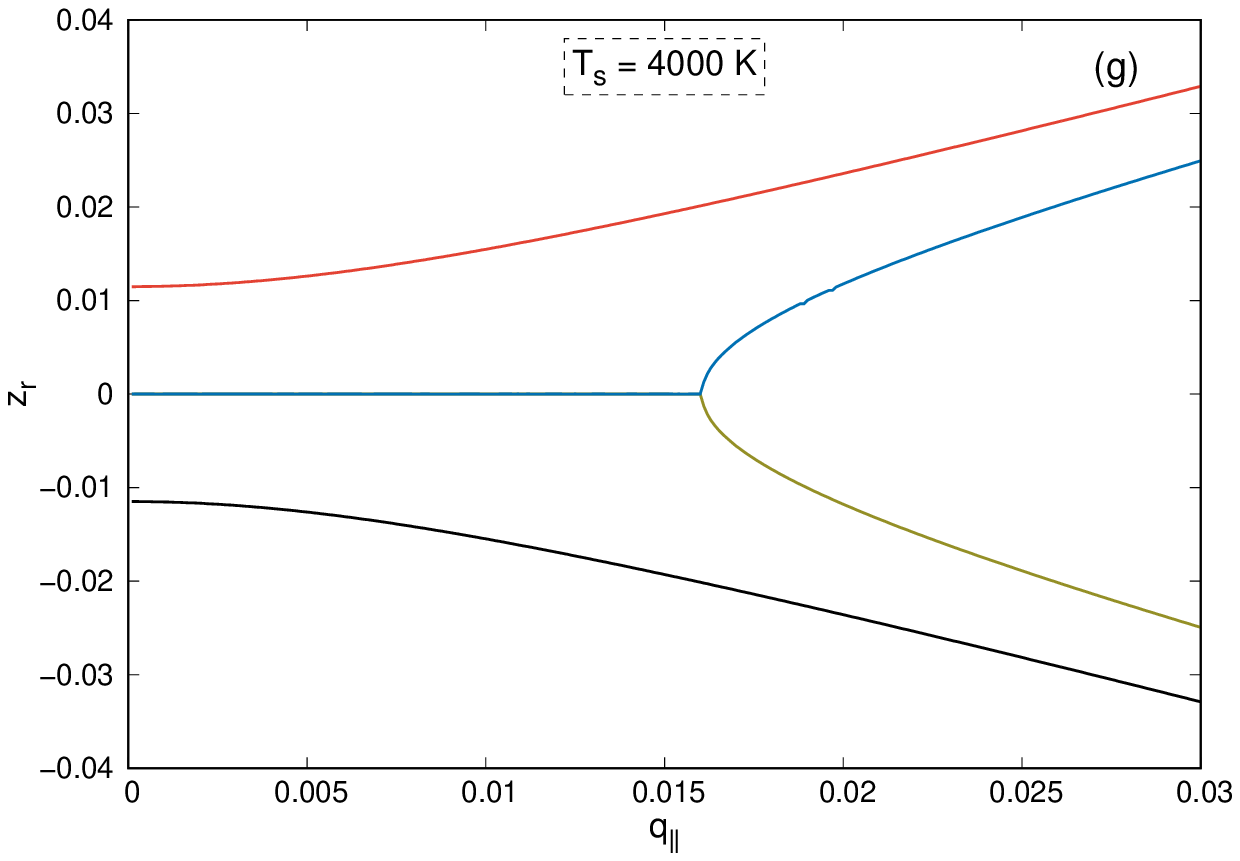}%
    \end{minipage}\hfill{}%
    \begin{minipage}[t]{0.49\textwidth}%
        \centering
        \includegraphics[width=.84\columnwidth]{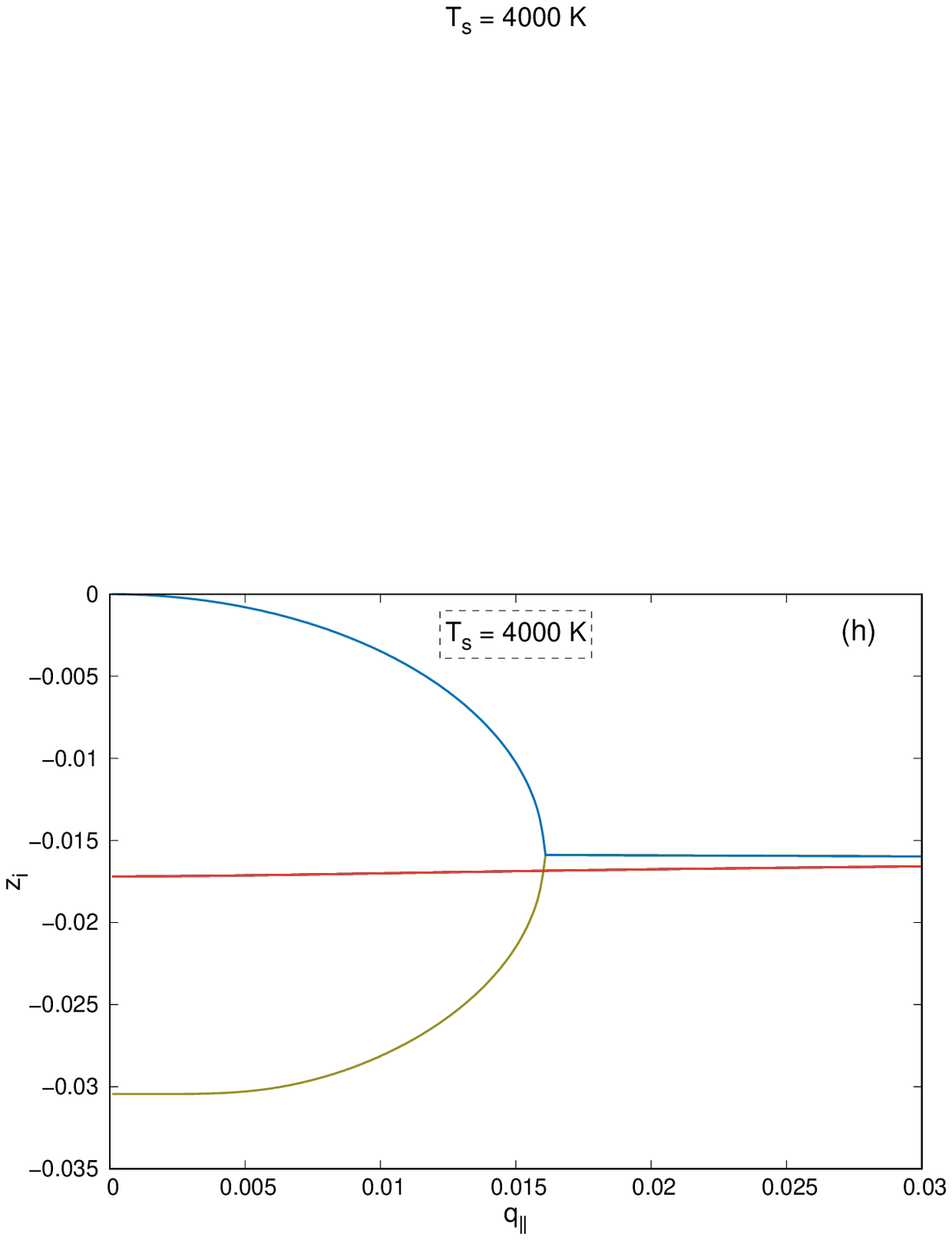}%
    \end{minipage}
    \caption{Real (left panels) and imaginary (right panels) parts of the normalized wave frequencies $z$ as a function of $q_\parallel$ with constant value of $q_\perp=0.02$ and $\varepsilon=2.5\times10^{-6}$ for four distinct values of radiation intensity, given by the stellar surface temperature $T_\mathrm{s}$. Red/black lines correspond to forward/backward propagating whistler modes, blue/yellow lines correspond to forward/backward ion cyclotron modes. Grey lines correspond to regions of no propagating waves, with both modes featuring zero value of $z_r$.}
    \label{fig:z_q_nonprop}
\end{figure*}

When we consider a non zero value of $T_s$, the photoionization process will alter the dust electrical charge, which becomes less negative, since dust particles will lose electrons by photoemission. Once the radiation flux is sufficiently strong, the dust particles will reach zero value of electrical charge and will start increasing for larger $T_s$ when $Z_\mathrm{d}>0$. This can be observed in Fig.~\ref{fig:z_q_Ts} where we show in its caption the numerical values for $Z_d$ in each case, e.g., when $T_s\simeq4618$\,K, which result in $Z_\mathrm{d}=0$, and when $T_s=5000$ and $5500$\,K, resulting in positively charged dust particles.

Fig.~\ref{fig:z_q_Ts} also shows that the photoionization process will also have effects on the propagation and damping of oblique Alfv{\'e}n waves. In panel (a) we see that the increase of radiation flux will diminish the non propagation region of the ion cyclotron modes. The imaginary part of $z$, depicted in panel (b), shows that the damping rate of the whistler and ion cyclotron modes will decrease with increasing stellar temperature for this set of parameters, but presenting similar behaviours for both negatively and positively charged dust grains. 

We point out that \citet{Gaelzer_2008} (fig. 3) have shown that the decrease of dust density will also reflect in a smaller non propagating region for ion cyclotron waves and smaller damping rates for all modes. This is expected since smaller dust densities will decrease the inelastic collision frequency (see equation~\eqref{eq:colision_freq}) hence decreasing the damping rates, since inelastic collisions is the process responsible for the appearance of this additional damping mechanism in Alfv{\'e}n waves in a magnetized dusty plasma \citep[][]{dejuli_2005}. What we observed in Fig.~\ref{fig:z_q_Ts} is that the increase of radiation flux and, consequently, the increase of dust electrical charge, will result in a similar behaviour of what is seen when dust density is decreased.

Another interesting feature presented in the work of \citet{Gaelzer_2008} is that for a given value of $q_\perp$ the obliquely propagating waves may show a region of $q_\parallel$ values where all modes are non propagating, this case is reproduced here in the top panels of Fig. \ref{fig:z_q_nonprop} for $T_s =0$\,K (no radiation flux). For this given set of parameters, we observe in panel (a) that in the region of very small $q_\parallel$ the whistler modes (red and black curves) present diminishing values of $z_r$ until $q_\parallel \approx 0.005$, at which point they become non propagating (as the ion cyclotron modes). This region is presented with grey curves and goes until $q_\parallel \approx 0.008$, after that the whistler modes become propagating again while the ion cyclotron modes (blue and yellow curves) will present non zero values of $z_r$ only for $q_\parallel \gtrsim 0.02$.

\begin{figure}
    \centering
    \includegraphics[width=\columnwidth]{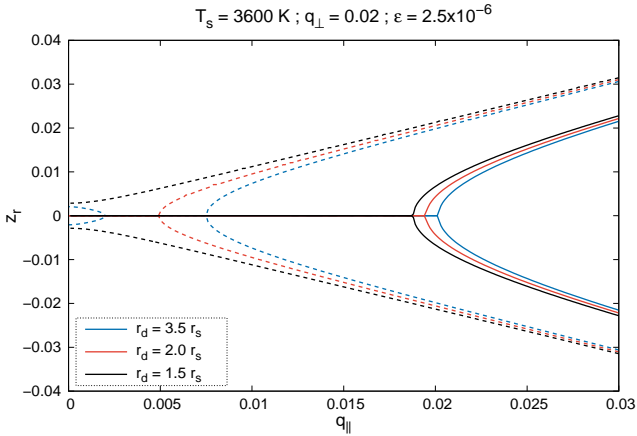}
    \caption{Real part of the normalized frequency $z_\mathrm{r}$ as a function of $q_\parallel$ for $q_\perp=0.02$, $\varepsilon=2.5\times10^{-6}$, $T_\mathrm{s}=3600$\,K and several values of mean dust distance $r_d$ as a multiple of stellar surface radius $r_s$. Dashed lines correspond to whistler modes while continuous lines correspond to ion cyclotron modes.}
    \label{fig:zr_q_rd}
\end{figure}

We see by the imaginary parts of $z$ in Fig. \ref{fig:z_q_nonprop}(b) that the forward and backward propagating whistler modes present the same damping rate in the regions where they have non zero values of $z_r$ (red line superposing black line). However, they will feature distinct values of $z_i$ in the region of non propagation (grey lines). In this region, the forward and backward ion cyclotron modes also present distinct values of $z_i$, and it makes no sense identifying any of these solutions since they all present $z_r=0$. The forward and backward ion cyclotron modes will also show distinct $z_i$ values as long as they present null $z_r$, and will have similar damping rates after $q_\parallel \simeq 0.02$ (blue line superposing yellow line) at which point they present non zero values of $z_r$.

Now we analyse the evolution of the mentioned regions (separated by black bars) for the case where photoionization of dust particles is considered. Figs. \ref{fig:z_q_nonprop}(c)-(h) consider a blackbody radiation flux coming from a stellar surface of temperatures $3400$, $3600$ and $4000$\,K. We notice that, considering the same set of parameters as in panels (a)-(b), the presence of a radiation flux will shrink the region where whistler waves present diminishing values of $z_r$ for increasing $q_\parallel$, until it eventually disappears in panels (e)-(f) for $T_s=3600$\,K.

The region where all modes are non propagating, delimited by black bars, will firstly increase with increasing temperature $T_s$ until the whistler modes no longer start with a non zero value of $z_r$ for $q_\parallel=0$. At this point we no longer see diminishing values of $z_r$ for whistler modes, and the non propagating region will shrink for greater values of $T_s$ until it eventually disappears, as in panels (g)-(h) for $T_s=4000$\,K. For even greater values of radiation flux we expect the ion cyclotron modes to present smaller region of zero $z_r$, and all modes will have smaller values of damping rates, similar of what we have seen in figure \ref{fig:z_q_Ts}\@.

A similar effect of what we just discussed is observed for a fixed stellar temperature $T_s$ and different distances $r_d$ of the dust grains from the star's radiating surface in equation~\eqref{eq:blackbody}\@. Fig.~\ref{fig:zr_q_rd} shows the case of $T_s=3600$\,K and same set of parameters as in Fig.~\ref{fig:z_q_nonprop} for distinct dust particles distance. The case represented by the red curves, where the mean dust distance is two times the radius of the star's radiating surface, is the same already seen in panel (e) of Fig.~\ref{fig:z_q_nonprop}\@. In this case, we see that whistler waves (dashed lines) present a region of zero $z_r$ for $q_\parallel \lesssim 0.005$, after which they become propagating.

We notice that for smaller distances such as $r_d/r_s=1.5$ (black lines) the whistler modes present non zero values of $z_r$ for all values of $q_\parallel$, as expected from what we learned in Fig.~\ref{fig:z_q_nonprop}, since dust particles are exposed to higher values of radiation flux. On the other hand, when we consider greater distances such as $r_d/r_s=3.5$ (blue lines) the radiation flux gets smaller and we can see a region of small $q_\parallel$ where whistler waves present decreasing values of $z_r$ and a region where all modes are non propagating. 

Therefore, it is possible to observe the distinct scenarios discussed in Fig.~\ref{fig:z_q_nonprop} also for a fixed value of stellar temperature (that is, for a given star) at different distances of the star's radiating surface, given that the plasma parameters remain the same. We remind the reader that, as a first approximation, equation~\eqref{eq:blackbody} does not consider the extinction of light, which could decrease even more the radiation flux as we go further from the stellar surface, since more dust particles would be along the way to absorb and scatter the radiation.

\section{Conclusions} \label{sec:conclusions}
Using a model for magnetized dusty plasmas within the kinetic theory, we have studied the characteristics of Alfv{\'e}n waves propagating obliquely to the direction of the ambient magnetic field in a stellar wind from a carbon-rich star. The employed formulation considers that the plasma species are described by a Maxwellian distribution of momenta and dust particles are immobile, given their much higher mass. Consequently, the studied wave frequencies are much larger than the usual dust frequencies, excluding modes that arise from dust dynamics. The dust particles are charged by absorption of plasma particles due to inelastic collisions and by photoionization. Our analysis focus on the effects caused by the photoionization process on the propagation and damping of kinetic Alfv{\'e}n waves (KAW), comparing with previous results where only absorption of particles is considered as dust charging mechanism.

We noticed that the presence of dust particles greatly modify both the real and imaginary parts of the normalized wave frequency of KAWs. For the considered parameters, shear Alfv{\'e}n waves always show a region of $q_\parallel$ values where the real part of the normalized frequency $z_r$ is zero. This interval of $q_\parallel$ values is modified when we consider the photoionization process, being diminished when the radiation flux incident on dust grains is increased. 

The damping rates, given by the imaginary parts of the normalized wave frequency $z_i$, of both compressional and shear Alfv{\'e}n modes also present distinct behaviours in a dusty plasma when compared to a conventional plasma. It was already known that dust grains cause the appearance of a new damping mechanism related to the inelastic collisions between dust grains and plasma particles. Consequently, shear Alfv{\'e}n waves, which present negligible damping rates in a conventional plasma, will show significant damping in a dusty plasma. The damping of compressional Alfv{\'e}n is specially different at small values of $q_\parallel$, where it presents practically constant values in a dusty plasma, unlike the complex behaviour presented for a conventional plasma. The incidence of radiation flux does not modify qualitatively the behaviour of these damping rates, at least for the parameters considered in this work, but will significantly diminish these values of $|z_i|$ for both modes if the radiation flux is sufficiently strong.

We also analysed the dispersion properties of KAWs for a particular set of parameters where both Alfv{\'e}nic modes present a region of $q_\parallel$ values with $z_r=0$. This case also shows the whistler/compressional Alfv{\'e}n waves with negative values of parallel group velocity for small $q_\parallel$. As we consider stars with greater stellar temperature $T_s$, we observe that the region where all modes are non propagating increases, until the whistler waves no longer present negative parallel group velocity. For even larger values of radiation flux, the region of non propagation starts to decrease until it vanishes for a sufficiently large value of stellar temperature. We also noticed that this effect can be observed in a given star (with constant stellar temperature) at different distances from the star's radiating surface, giving that the plasma parameters remain the same.

These results show that the presence of dust particles significantly modify the dispersion characteristics and absorption of KAWs, and that the photoionization process may be of great importance in situations where it dominates over other charging mechanisms, altering the equilibrium electrical charge of dust particles. Such situations can be found in the winds of a relatively hot carbon-rich star, although this formalism can be applied for other space environments, such as stellar winds of stars from a different category or the magnetosphere of planets, bearing in mind that other charging mechanisms may bring greater effects than the photoionization in systems with hotter and denser plasma.

It is possible to see in the figures presented in this work some features of the group velocity of the studied Alfv{\'e}nic modes. We can see by the slopes of the curves that, in the presence of dust particles, the ion cyclotron modes will always show a region of $q_\parallel$ values with zero parallel component of the group velocity, beginning to show positive (negative) values when the forward (backward) propagating mode shows non zero $z_r$. Meanwhile, a specific whistler mode could present in a given situation (see e.g. Fig.~\ref{fig:z_q_nonprop}(a)) negative, null or positive parallel group velocity for distinct regions of $q_\parallel$ values. However, this study does not give any information regarding the perpendicular component of group velocity. This motivate us to pursuit a better understanding of the evolution of both parallel and perpendicular components of group velocities for KAWs, since this property will provide information regarding the energy propagation of these waves. We are currently working in this subject and intend to publish our results in the near future.

Several other avenues of investigation can be followed from this work. The model has several limitations and could be greatly improved. As already mentioned, the model assumes a single grain size, which is convenient to derive a relatively simple dispersion relation to work with, but such assumption is somewhat unrealistic, given that space environments present several dust populations of different sizes. \citet{galvao_ziebell2012} have derived an expression of the dielectric tensor for a magnetized plasma considering that dust particles present a discrete distribution of sizes, charged by inelastic collisions and by photoionization. We intend to make use of this formulation to include a continuous distribution function of grain sizes to the model and investigate the consequences to the waves' properties.

The expression for the radiation flux, given by equation~\eqref{eq:blackbody}, can be improved by considering the effect of extinction. This would require a knowledge about the distribution of dust particles between the source of radiation and the observer, and is to a certain degree related to the problem mentioned above regarding the distribution of dust sizes in the star's vicinity. The model of light extinction used in the work of \citet{truong2021modeling} separates the circumstellar dust shell of the star Betelgeuse in several layers with distinct plasma and dust properties, showing results that well reproduce the observed flux from the near-UV to near-IR range. This model, which can provide the reddening of light in a region within the dust shell, could be well suited to future studies using the formulation presented here.

The formalism applied is based on the linearized from of the Klimontovich-Maxwell set of equations and does not take into account higher non-linear terms of this system. Thus, a more comprehensive treatment is necessary to better understand the several non-linear nature of the local wave-particle and wave-wave interactions. The results seen here and in several other works indicate that the dust population will ultimately affect the non-linear processes where Alfv{\'e}n waves are involved in the dust-rich environment of carbon-rich stars. We also intend to pursuit this line of investigation.

\section*{Acknowledgements}

This study was financed in part by the Coordena{\c{c}}{\~a}o de Aperfei{\c{c}}oamento de Pessoal de N{\'i}vel Superior – Brasil (CAPES) – Finance Code 001. RG acknowledges support from CNPq (Brazil), grant No. 307845/2018-4. LFZ acknowledges support from CNPq (Brazil), grant No. 302708/2018-9.

%%%%%%%%%%%%%%%%%%%%%%%%%%%%%%%%%%%%%%%%%%%%%%%%%%
\section*{Data Availability}

The data underlying this article will be shared on reasonable request to the corresponding author.

%%%%%%%%%%%%%%%%%%%% REFERENCES %%%%%%%%%%%%%%%%%%

% The best way to enter references is to use BibTeX:

\bibliographystyle{mnras}
\bibliography{MAIN.bib} % if your bibtex file is called example.bib

% Alternatively you could enter them by hand, like this:
% This method is tedious and prone to error if you have lots of references
%\begin{thebibliography}{99}
%\bibitem[\protect\citeauthoryear{Author}{2012}]{Author2012}
%Author A.~N., 2013, Journal of Improbable Astronomy, 1, 1
%\bibitem[\protect\citeauthoryear{Others}{2013}]{Others2013}
%Others S., 2012, Journal of Interesting Stuff, 17, 198
%\end{thebibliography}

%%%%%%%%%%%%%%%%%%%%%%%%%%%%%%%%%%%%%%%%%%%%%%%%%%

%%%%%%%%%%%%%%%%% APPENDICES %%%%%%%%%%%%%%%%%%%%%

%%%%%%%%%%%%%%%%%%%%%%%%%%%%%%%%%%%%%%%%%%%%%%%%%%

% Don't change these lines
\bsp	% typesetting comment
\label{lastpage}
\end{document}